\documentclass[12pt,twocolumn]{iopart}
\usepackage{iopams}
\usepackage{graphicx}
\usepackage[usenames]{color}
\usepackage[normalem]{ulem}
\begin{document}
\title{Spatiotemporal Binary Interaction and Designer quasi particle Condensates}
\author{Ramaswamy Radha$^1$, Pattu Sakthi Vinayagam$^1$,Hyun Jong Shin$^2$ and Kuppuswamy Porsezian$^{3}$}
\address{$^1$ Centre for Nonlinear Science, PG and Research Dept. of Physics, Govt. College
for Women (Autonomous), Kumbakonam - 612001, India. \\
  $^2$Department of Physics and Research Institute of Basic Sciences, Kyung Hee University, Seoul 130-701, Korea \\
  $^3$Department of Physics, Pondicherry University,
Pondicherry-605014, India}
 \ead{radha\_ramaswamy@yahoo.com $^{1}$}
 \ead{hjshin@khu.ac.kr$^{2}$}
 \ead{ponzsol@yahoo.com$^{3}$}

\begin{abstract}
We introduce a new integrable model to investigate the dynamics of
two component quasi particle condensates with spatio temporal
interaction strengths. We derive the associated Lax-pair of the
coupled GP equation and construct matter wave solitons. We show
that the spatio temporal binary interaction strengths not only
facilitate the stabilization of the condensates, but also enables
one to fabricate condensates with desirable densities, geometries
and properties leading to the so called "designer quasi particle
condensates".
\end{abstract}

\hspace{18mm}{\bf Keywords:}Gauge transformation; Bright
soliton;GP equation.


 \maketitle
\section{Introduction}
The experimental observation of Bose-Einstein condensates (BEC) in
rubidium by Cornell and Wieman [1] and in sodium by Ketterle's
group [2] have signalled a new era in ultra cold atomic physics
and have virtually opened the flood gates for matter wave
manipulation. At ultra low temperatures, the behaviour of BECs is
governed by the Gross-Pitaevskii (GP) equation which is nothing
but the inhomogeneous (3+1) dimensional nonlinear Schr\"odinger
(NLS) equation of the following form [3-6]
\begin{equation}
i\hbar \frac{\partial \psi(\vec{r},t)}{\partial t} =
\left(-\frac{\hbar^2}{2 m}\nabla^2+U|{\psi(\vec{r},t)}|^2
+V(\vec{r},t) \right){\psi(\vec{r},t)},
\end{equation}
where $ U(\vec{r},t) $ describes the interatomic interaction and $
V(\vec{r},t) $ represents the external trapping potential that
traps the atoms at such low temperatures. Eventhough the above
(3+1) dimensional GP equation is in general nonintegrable, it has
been shown to admit integrability in quasi one dimensions for both
time independent [7,8] and time dependent [9,10] harmonic trapping
potentials. The experimental observation of bright [11-13] and
dark solitons [14,15] in such quasi one dimensional BECs has not
only reconfirmed the integrability  of the associated dynamical
systems,  but also stimulated  a lot of interest to get a deeper
understanding of the nonlinear phenomena surrounding BECs.\\

 $\quad$ It must be mentioned that the behaviour of single component
  BECs is controlled by the external trapping potential
 and the binary interatomic interaction (scattering
 length). In contrast to single component BECs, the dynamics of
 multicomponent BECs [16-19] which comprise of either mixtures of
 different hyperfine states of the same atomic species or even mixtures
  of different atomic species is much richer
 by virtue of both interspecies interaction and intraspecies interaction.
 Multicomponent BECs show several interesting and novel phenomena
 like soliton trains, multidomain walls, spin switching [20],
  multi mode collective excitations etc., which are not normally
 encountered in single component BECs. The recent analytical
 investigations of multicomponent BECs [21,22] involving temporal
 variation of both interspecies and intraspecies scattering
 lengths have shown how the concept of coherent storage and matter
 wave switching can be manifested in the collisional dynamics of
 bright solitons.

$\quad$ It is worth pointing at this juncture that the above
scalar and vector BECs involve the condensation of atoms or
particles with integral spin (bosons) at the ground state. Since
the temperature at which bosons condense is governed by the
inverse of the mass, the low density of the weekly interacting
atoms combined with their relatively large mass  ensures that the
critical temperature is extremely low of the order order $10^{-6}
k$. In this context, it was believed that the condensation of
quasi particles like excitons and polaritons with their high
densities and negligible masses might occur at temperatures of
several orders of magnitude higher than for atoms, reachable by
standard cryogenic techniques. In this connection, the advent of
exciton BEC [23] and a polariton BEC (or) polariton laser[24] have
certainly contributed to a resurgence in this exotic state of
matter and have heralded a new era in semiconductor
heterostructures. Excitons and polaritons are endowed with the
spatially varying masses which ensures that the inter
quasi-particle interaction could be spatially inhomogeneous.
Recently, such quasi particle condensates were modelled [25] by
the Gross- Pitaeveskii equation with a space dependent dispersion
coefficient reperesenting the position dependent masses of quasi
particles of the form
\begin{equation}
i q_t + k(x)q_{xx}+a(x)|q|^2 q +  v(x)q =0
\end{equation}
In the above equation, the dispersion coefficient $k(x)$, the
scattering length $a(x)$ and the trap $v(x)$ are related by the
following equations
\begin{equation}
k(x) = \frac {4}{a(x)^2}
\end{equation}
and
\begin{equation}
v(x)=\frac{-3 a'(x)^2+2 a(x) a''(x)}{a(x)^4}.
\end{equation}

$\quad$It should be mentioned that eq.(2) has been mapped on to
the well known NLS equation and has been investigated earlier
[26-31]. In this context, it would be interesting to investigate
the dynamics of two component quasi particle condensates. The fact
that the two component quasi particle condensates are endowed with
both intraspecies interaction and interspecies interaction means
that one could conceive of spatio temporal binary interaction in
them.

$\quad$Motivated by the above consideration, we investigate the
dynamics of two component quasi particle condensates with space
and time modulated nonlinearities. In particular, we derive the
lax pair of the associated Gross-Pitaeveskii equation and generate
bright solitons. We also study the collisional dynamics of matter
wave solitons in harmonic and optical lattice potentials.

\section{Model and Lax-pair}
$\quad$Considering a  temporally and spatially inhomogeneous two
component quasi particle BEC, the behaviour of the condensates
that are prepared in two hyperfine states of the same atom can be
described by the two coupled GP equation of the following form

\begin{eqnarray}
 i\psi_1 (x,t)_t &+& k(x)\psi_1 (x,t)_{xx} + \frac{2}{\sqrt
 {k(x)}}(A_1 (t)^{2}|\psi_1 (x,t)|^{2} \nonumber \\
 &+& A_2 (t)^{2}|\psi_2 (x,t)|^{2})\psi_1 (x,t) +(V(x,t)+i \frac{A_1 (t)_{t}}{A_1 (t)}\nonumber\\
 &&-\frac{i}{2} {b_1 (t)}-\theta_1(t) _t)\psi_1 (x,t)=0,
 \end{eqnarray}
 \begin{eqnarray}
 i\psi_2 (x,t)_t &+& k(x)\psi_2 (x,t)_{xx} + \frac{2} {\sqrt
 {k(x)}}(A_1 (t)^{2}| \psi_1 (x,t)|^{2} \nonumber \\
 &+& A_2 (t)^{2}|\psi_2 (x,t)|^{2})\psi_2 (x,t)+(V(x,t)+i \frac{A_2 (t)_{t}}{A_2 (t)}\nonumber\\
 &&-\frac{i}{2} {b_1 (t)}-\theta_2(t) _t) \psi_2 (x,t)=0,
\end{eqnarray}
where
\begin{eqnarray}
V(x,t)= \frac{3}{16}\frac{k_{x}^{2}}{k} -\frac{1}{4}k_{xx}+
\frac{1}{4} K^2 {b_1}^{2} -\frac{1}{4} K^2 {b_1}_{t}- {b_0} K,
\end{eqnarray}
and
\begin{equation}
K(x)=\int \frac{1}{{\sqrt {k(x)}}} dx.
\end{equation}
In the above equation, $A_1 (t),A_2 (t),b_0 (t), b_1 (t), \theta_1
(t)$ and $\theta_2 (t)$ are arbitrary functions of time. The
arbitrary functions $A_1 (t)$and $A_2 (t)$ describe the temporal
variation of binary interaction while the arbitrary functions
($b_0 (t), b_1 (t)$) and ($\theta_1 (t),\theta_2 (t)$) facilitate
us to choose the potential and atomic feeding respectively.  In
the above equation, the condensate wave functions are normalized
to particle numbers $N_i = \int |\psi_{i}|^{2} d^{3}r$ while
$k(x)$ which represents the spatially varying dispersion
coefficient is also related to the mass of the quasi-particles.
Equations (5) and (6) admit the following Lax-pair
\begin{eqnarray}
\Phi_x &+& U \Phi=0,\\
\Phi_t &+& V \Phi=0,
\end{eqnarray}

where $\Phi = (\phi_1, \phi_2, \phi_3)^T$ and
\begin{eqnarray}
U &=& \left(%
\begin{array}{ccc}
\frac{i}{2}\lambda & \frac{\psi_1}{k^{3/4}} &
\frac{\psi_2}{k)^{3/4}}\\
-\frac{\psi_1^*}{k^{3/4}}& -\frac{i}{2}\lambda & 0 \\
-\frac{\psi_2^*}{k^{3/4}}& 0 & -\frac{i}{2}\lambda\\
\end{array}%
\right),
\end{eqnarray}
\begin{eqnarray}
V&=&\left(%
\begin{array}{ccc}
V_{11} & V_{12}& V_{13}\\
V_{21}& V_{22}& V_{23}\\
V_{31}& V_{32} & V_{33}\\
\end{array}%
\right),
\end{eqnarray}
with
\begin{eqnarray}
V_{11} &=& -i \frac{\psi_1 \psi_1^*}{{\sqrt {k}}}-i\frac{\psi_2
\psi_2^*}{{\sqrt {k}}}+ \frac{i}{2}\lambda^{2}k+i g; V_{12} =
\chi_1+\lambda k^{1/4}\psi_1, V_{13}=\chi_2+\lambda
k^{1/4}\psi_2,\nonumber \\
V_{21} &=& -\chi_1^*-\lambda k^{1/4}\psi_1^*, V_{22} =
i\frac{\psi_1 \psi_1^*}{{\sqrt {k}}}- \frac{i}{2}\lambda^{2}k-i g;
V_{23} = i \frac{\psi_2 \psi_1^*}{{\sqrt {k}}},\nonumber \\
V_{31} &=& -\chi_2^*-\lambda k^{1/4}\psi_2^*; V_{32} =
i\frac{\psi_2^* \psi_1}{{\sqrt {k}}}; V_{33} = i \frac{\psi_2
\psi_2^*}{{\sqrt {k}}}- \frac{i}{2}\lambda^{2}k-i g,\nonumber
\end{eqnarray}
\begin{eqnarray}
\chi_1&=& \frac {- \frac{i}{4}\psi_1 k_x}{k^{3/4}} + i k^{1/4}
\psi_{1x}, \chi_1 ^{*}= \frac{\frac{i}{4}\psi_1 k_x}{k^{3/4}}-
i^{1/4}\psi_{1x}^{*},\nonumber\\
\chi_2&=& \frac {- \frac{i}{4}\psi_2 k_x}{k^{3/4}} + i k^{1/4}
\psi_{2x}, \chi_2 ^{*}= \frac{\frac{i}{4}\psi_2 k_x}{k^{3/4}}- i
k^{1/4}\psi_{2x} ^{*},\nonumber\\
g(x)&=& {1 \over 8} (b_{1t} -b_1^2 ) K^2 + {1 \over 2} b_0 K.
\end{eqnarray}
In the above equation,
$\lambda(x,t)=\frac{b_1(t)K(x)/2+\Lambda(t)}{\sqrt {k(x)}}$
represents the complex nonisospectral parameter and $\Lambda(t)=
\int {b_0(t)} e^{(-\int {b_1 (t)}dt) dt + \mu)} e^{(\int {b_1
(t)}dt)}$ is a complex function of time while  $\mu$ is the so
called " hidden  complex spectral parameter ". The temporal
scattering lengths have been absorbed into $\psi_1 (x,t)$ and
$\psi_2(x,t)$ by substituting $\psi_1 (x,t) \rightarrow A_1
(t)e^{i \theta_1 (t)} \hat \psi_1 (x,t)$ and $\psi_2 (x,t)
\rightarrow A_1 (t)e^{i \theta_1 (t)} \hat \psi_2 (x,t)$. Under
the following transformation
\begin{eqnarray}
q_1(X, T) &=& A_1 (t) {\rm exp}\left(i \theta_1 (t) -i \Pi \right) k(x)^{-1/4} I^{-1} \psi _1 (x,t) , \nonumber \\
q_2(X, T) &=& A_2 (t) {\rm exp} \left(i \theta_2 (t) -i \Pi
\right) k(x)^{-1/4} I^{-1} \psi _2 (x,t) ,
\end{eqnarray}
where
\begin{eqnarray}
X &=& I K(x) +2 \int I^2 J dt, \nonumber \\
T &=& \int I^2 dt, \nonumber \\
I &=& {\rm exp} \left(\int b_1 (t) dt \right), \nonumber \\
J &=& \int b_0 (t) {\rm exp} \left(-\int b_1 (t) dt \right) dt, \nonumber \\
\Pi &=& -I J K(x) - {1 \over 4} b_1 (t) K(x)^2 - \int I^2 J^2
dt,\nonumber
\end{eqnarray}
eqs.(5) and (6) transform to the standard 2-component NLS equation
[32].

\section{Construction of Bright Vector  Solitons}
    To generate the bright vector solitons of the coupled GP equations  $(5)$ and $(6)$,
we now consider the vacuum solution ($\psi_1^{0} =
\psi_2^{(0)}=0$) so that the corresponding eigenvalue problem
becomes
\begin{eqnarray}
\Phi_x^{(0)} + U^{(0)} \Phi^{(0)}=0,\\
\Phi_t^{(0)} + V^{(0)} \Phi^{(0)}=0,
\end{eqnarray}
where
\begin{equation}
U^{(0)}=\left(%
\begin{array}{ccc}
\frac{i}{2}\lambda(x,t) & 0 &
0\\
0& -\frac{i}{2}\lambda(x,t) & 0 \\
0& 0 & -\frac{i}{2}\lambda(x,t)\\
\end{array}%
\right),
\end{equation}
\begin{equation}
V^{(0)}=\left(%
\begin{array}{ccc}
\frac{i}{2}\lambda^{2}k(x) & 0 & 0 \\
0& - \frac{i}{2}\lambda^{2}k(x)& 0 \nonumber \\
0& 0 & - \frac{i}{2}\lambda^{2}k(x)
\end{array}%
\right).
\end{equation}
Solving the above eigenvalue problem, one obtains the following
vacuum eigen function
\begin{equation}
\Phi^{(0)}=\left(%
\begin{array}{ccc}
\phi^{(0)11} & 0 & 0 \\
0 & \phi^{(0)22}  & 0 \\
0 & 0 &\phi^{(0)33} \\
\end{array}%
\right),
\end{equation}
where
\begin{eqnarray}
\phi^{(0)11}&=&e^{-\frac{i}{2}(b_1 (t) K(x)^2 /4 +\Lambda(t)K(x)+\int\Lambda(t)^{2} dt)},\nonumber\\
\phi^{(0)22}&=&e^{\frac{i}{2}(b_1 (t) K(x)^2 /4 +\Lambda(t)K(x)+\int\Lambda(t)^{2} dt)},\nonumber\\
\phi^{(0)33}&=&e^{\frac{i}{2}(b_1 (t) K(x)^2 /4
+\Lambda(t)K(x)+\int\Lambda(t)^{2} dt)}.\nonumber
\end{eqnarray}
We now gauge transform the vacuum eigenfunction $\Phi^{(0)}$ by a
transformation function $g(x,t)$ to give
\begin{eqnarray}
U^{(1)} = g U^{(0)} g^{-1}+g_x g^{-1}, \\
V^{(1)}= g V^{(0)} g^{-1}+g_t g^{-1}.
\end{eqnarray}
We now choose the transformation function $g(x,t)$ from the
solution of the associated Riemann problem such that it is
meromorphic in the complex $\lambda$ plane as
\begin{equation}
g(x,t;\lambda)=\left[1+\frac{\lambda_1-\bar{\lambda}_1}{\lambda-\lambda_1}P(x,t)
\right]\cdot\left(%
\begin{array}{ccc}
  1 & 0 & 0 \\
  0 & -1 & 0 \\
  0 & 0 & -1 \\
\end{array}%
\right).
\end{equation}
The inverse of matrix $g$ is given by
\begin{equation}
g^{-1}(x,t;\lambda)=\left(%
\begin{array}{ccc}
  1 & 0 & 0 \\
  0 & -1 & 0 \\
  0 & 0 & -1 \\
\end{array}%
\right)\cdot\left[1-\frac{\lambda_1-\bar{\lambda}_1}{\lambda-\bar{\lambda}_1}P(x,t)
\right]
\end{equation}
where $\lambda_1$ and $\bar{\lambda}_1 = \lambda_1^*$ are
arbitrary complex parameters and $P$ is a 3$\times$3 projection
matrix ($P^2 =P$) which can be obtained using vacuum eigen
function $\phi^{0}(x,t)$ as [33]
\begin{equation}
P=J \cdot \tilde{P}\cdot J,
\end{equation}
where
\begin{equation}
\tilde{P}= \frac{M^{(1)}}{\rm{Trace} [M^{(1)}]},
\end{equation}
\begin{equation}
J=\left(%
\begin{array}{ccc}
  1 & 0 & 0 \\
  0 & -1 & 0 \\
  0 & 0 & -1 \\
\end{array}%
\right).
\end{equation}
and
\begin{equation}
M^{(1)}=\Phi^{(0)}(x,t,\bar{\lambda}_1)\cdot\hat{m}^{(1)}\cdot\Phi^{(0)}
(x,t,\lambda_1)^{-1}.
\end{equation}
In the above equation, $\hat{m}^{(1)}$ is a 3$\times$3 arbitrary
matrix taking the following form
\begin{equation}
\hat{m}^{(1)}=\left(%
\begin{array}{ccc}
  e^{2\delta_1}\sqrt{2} & \varepsilon_1^{(1)}e^{2i\eta_1} & \varepsilon_2^{(1)}e^{2i\eta_1} \\
  \varepsilon_1^{*(1)}e^{-2i\eta_1} & e^{-2\delta_1}/\sqrt{2} & 0 \\
  \varepsilon_2^{*(1)}e^{-2i\eta_1} & 0 & e^{-2\delta_1}/\sqrt{2} \\
\end{array}
\right),
\end{equation}
such that the determinant $M^{(1)}$ becomes zero with the
condition $|\varepsilon_1^{(1)}|^2+|\varepsilon_2^{(1)}|^2
$=$cos(\phi)^{2}+ sin(\phi)^{2}$=1. Thus, choosing $\lambda_1 =
\alpha_1 +i\beta_1$ and $\bar{\lambda}_1 = \lambda_1^*$ and using
eq. (26), the matrix $M^{(1)}$ can be explicitly written as
\begin{equation}
M^{(1)}=\left(%
\begin{array}{ccc}
  e^{-\Theta_1}\sqrt{2} & e^{-i\xi_1}\varepsilon_1^{(1)} & e^{-i\xi_1}\varepsilon_2^{(1)} \\
  e^{i\xi_1} \varepsilon_1^{*(1)}& e^{\Theta_1}/\sqrt{2} & 0 \\
  e^{i\xi_1} \varepsilon_2^{*(1)} & 0 & e^{\Theta_1}/\sqrt{2} \\
\end{array}%
\right),
\end{equation}
where
\begin{eqnarray}
\Theta_1 &=& -\beta_{1}(t) K(x)- 2 \int \alpha_{1}(t) \beta_{1}(t) dt-2 \delta_1 ,\\
\xi_1 &=& -b_1 (t) K(x)^2 /4 -\alpha_{1}(t) K(x) \nonumber \\
&&-\int (\alpha_{1}(t)^{2}-\beta_{1}(t))^{2} dt -2\eta_1,
\end{eqnarray}
 with
 \begin{eqnarray}
 \alpha_1(t)&=&\left(\int b_0 (t) exp\left(-\int b_1 (t) dt \right) dt +\alpha_1 \right)\times exp \left({\int b_1 (t) dt}\right) , \nonumber \\
\beta_1(t)&=&\beta_1 exp \left({\int b_1 (t)  dt} \right),
\end{eqnarray}
while $\delta_1$ and $\eta_1$ are arbitrary parameters. Now,
substituting eqns.(17), (21) in eqn.(19), we obtain
\begin{eqnarray}
U^{(1)} = U^{(0)}
-2i(\lambda_1 - \bar{\lambda}_1)\left(%
\begin{array}{ccc}
  0 & \tilde{P}_{12} & \tilde{P}_{13} \\
 -\tilde{P}_{12} & 0 & 0 \\
  -\tilde{P}_{13} & 0 & 0 \\
\end{array}%
\right),
\end{eqnarray}

and similarly for $V^{(1)}$. Thus, one can write down the one
soliton solution as

\begin{equation}
U^{(1)}_{12} = U^{(0)}_{12} - 2i (\lambda_1 -\bar{\lambda}_1)
\tilde{P}_{12},
\end{equation}
\begin{equation}
U^{(1)}_{13} = U^{(0)} _{13}- 2i (\lambda_1 -\bar{\lambda}_1)
\tilde{P}_{13}.
\end{equation}
Thus, the explicit forms of bright soliton solution can be written
as

\begin{equation}
\psi_{1}^{(1)}= -cos(\phi)\beta_1(t) k(x)^{1/4}
sech(\Theta_1)e^{(i \xi_{1})},
\end{equation}
\begin{equation}
 \psi_{2}^{(1)}=
-sin(\phi)\beta_1(t) k(x)^{1/4} sech(\Theta_1)e^{(i \xi_{1})}.
\end{equation}
where $\phi$ is the arbitrary parameter. Looking at the above
solution for $\psi_{1}^{(1)}$ and $\psi_{2}^{(1)}$, we infer that
the amplitude of the bright solitons representing the condensates
can be spatially and temporally modulated. This means that one can
desirably change the intensities of the matter wave solitons by
choosing spatial and temporal inhomogeneities suitably. In
otherwords, spatially and temporally modulated  scattering lengths
can lead to various interesting profiles of the condensates.\nonumber\\

\section{Matter wave solitons and their interaction}

\subsection{Grating solitons in optical lattice potentials}
Choosing the dispersion coefficient $k(x)=(sin(0.1 x)^{2}- 4$ x
$0.5)^{4}$ x $0.3$, one obtains an optical lattice potential as
shown in figure (1a) (for an appropriate choice of the parameters
$\theta_1 (t), \theta_2 (t), {b_0 (t)}$ and ${b_1 (t)}$. Under
this condition, we observe that the matter waves are confined in
space periodically as shown in figs (1b) and (1c) and we call them
as "Grating solitons"[34]. The width and amplitude of the grating
solitons can be modulated by suitably changing the parameters
associated with $k(x)$.

\begin{figure}
\includegraphics[scale=0.4]{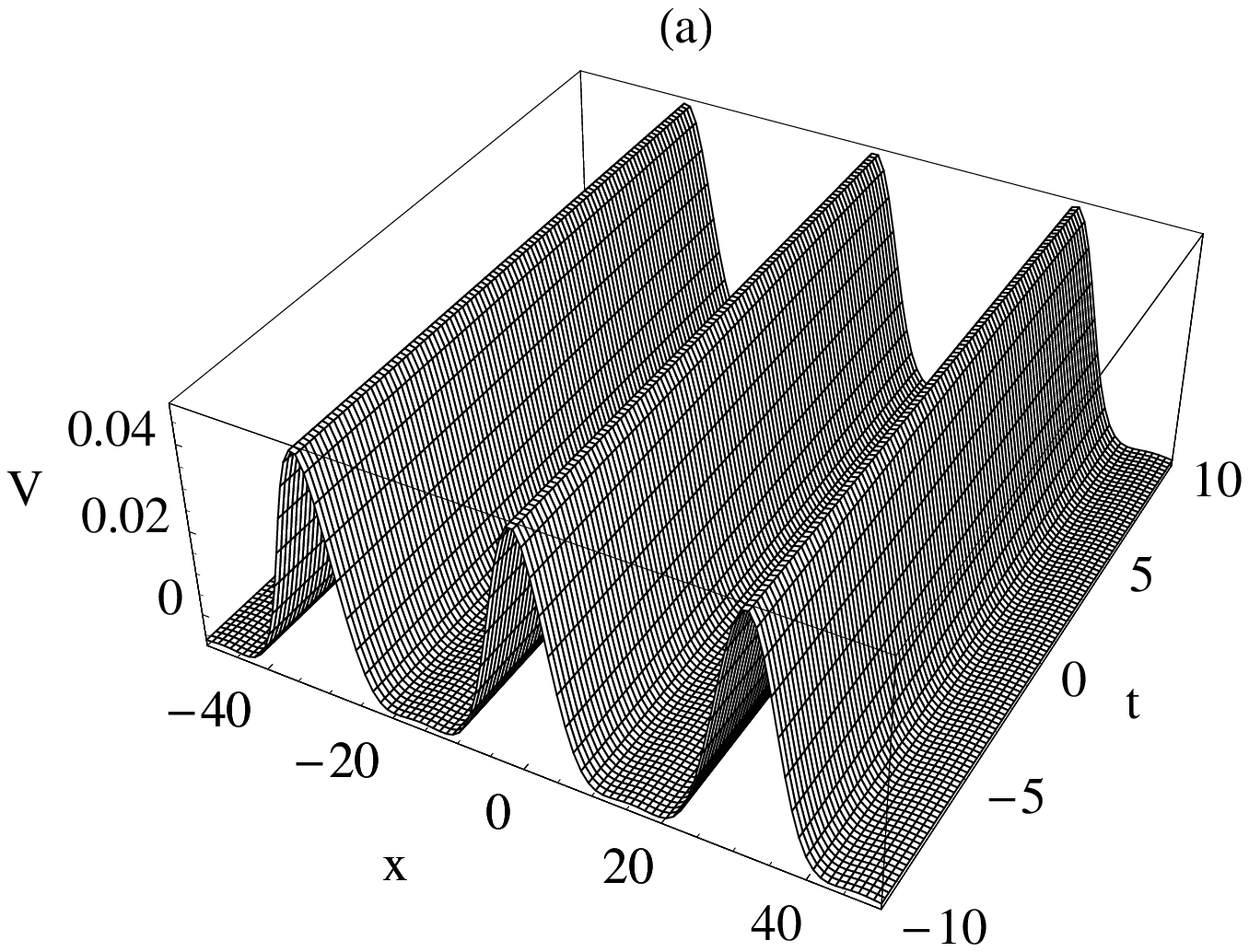}\includegraphics[scale=0.4]{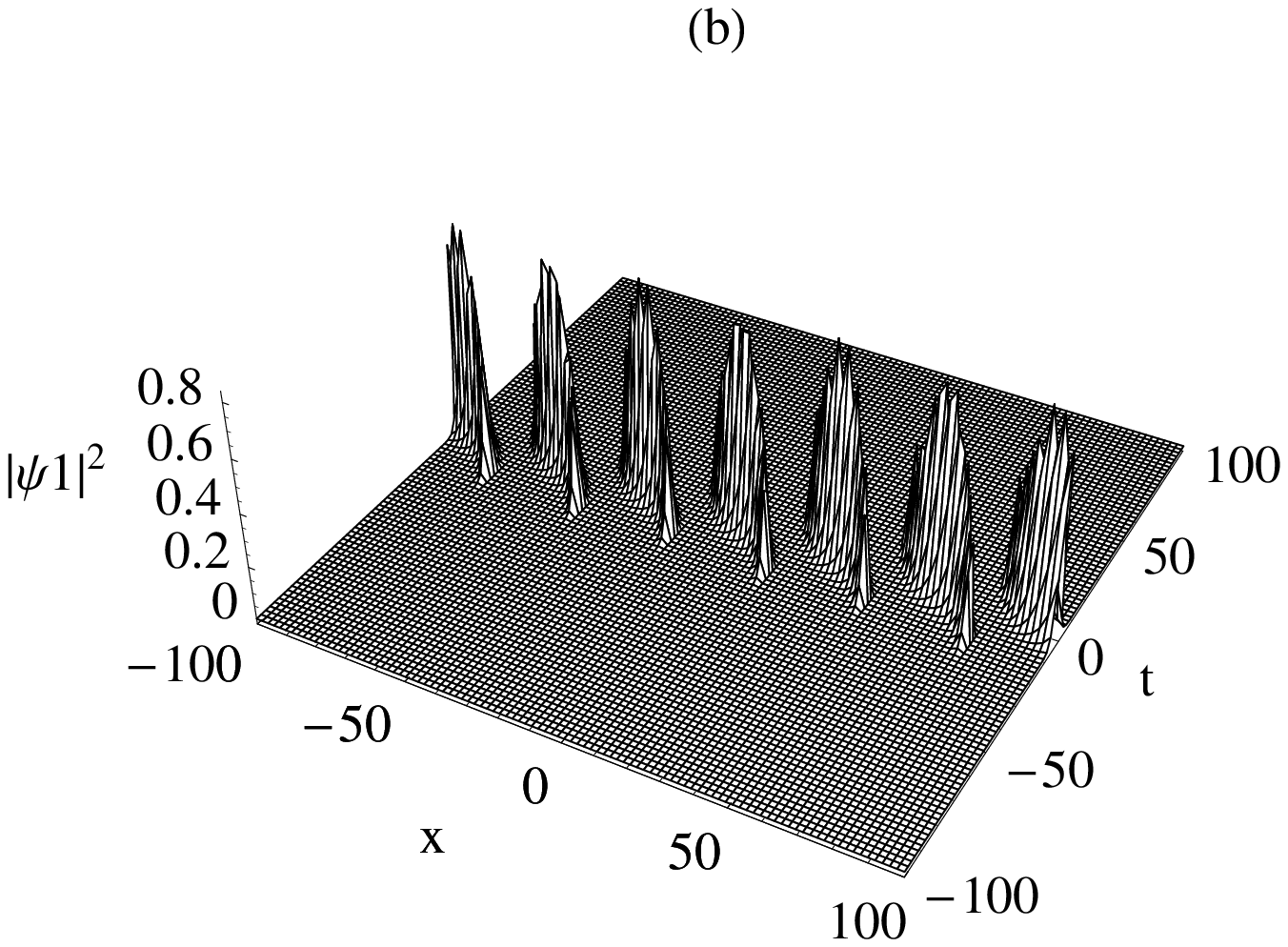}\includegraphics[scale=0.4]{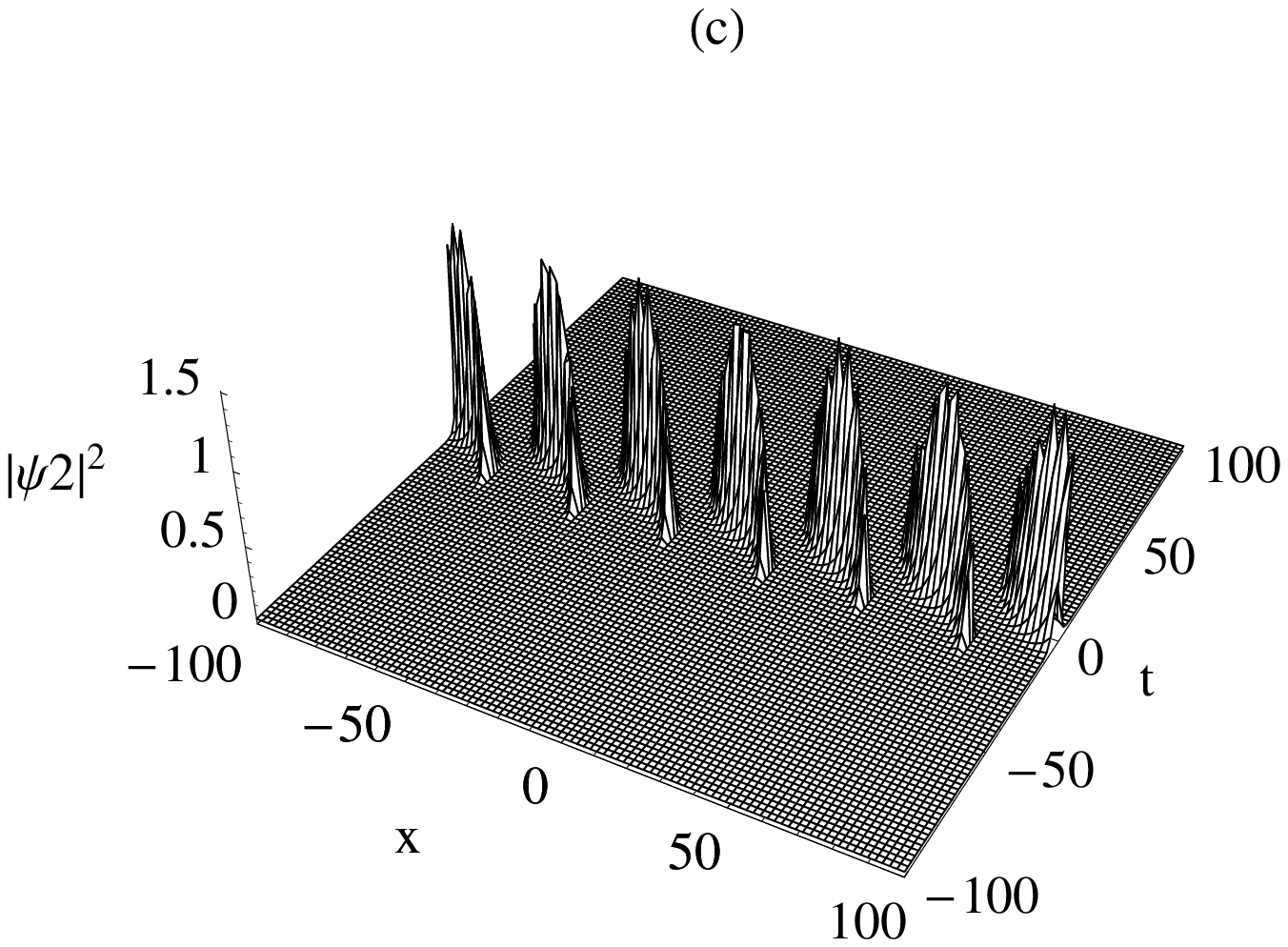}
\caption{(a) Optical lattice for the choice $k(x)=(sin(0.1 x)^{2}-
4$ x $0.5)^{4}$ x $0.3$ with $A_1 (t)=A_2 (t)=1,{b_0 (t)}={b_1
(t)}=0 $ and $ \theta _1(t)=\theta _2(t)=1 $ (b),(c) Grating
solitons in the optical lattice potentials.}
\end{figure}

\begin{figure}
\includegraphics[scale=0.4]{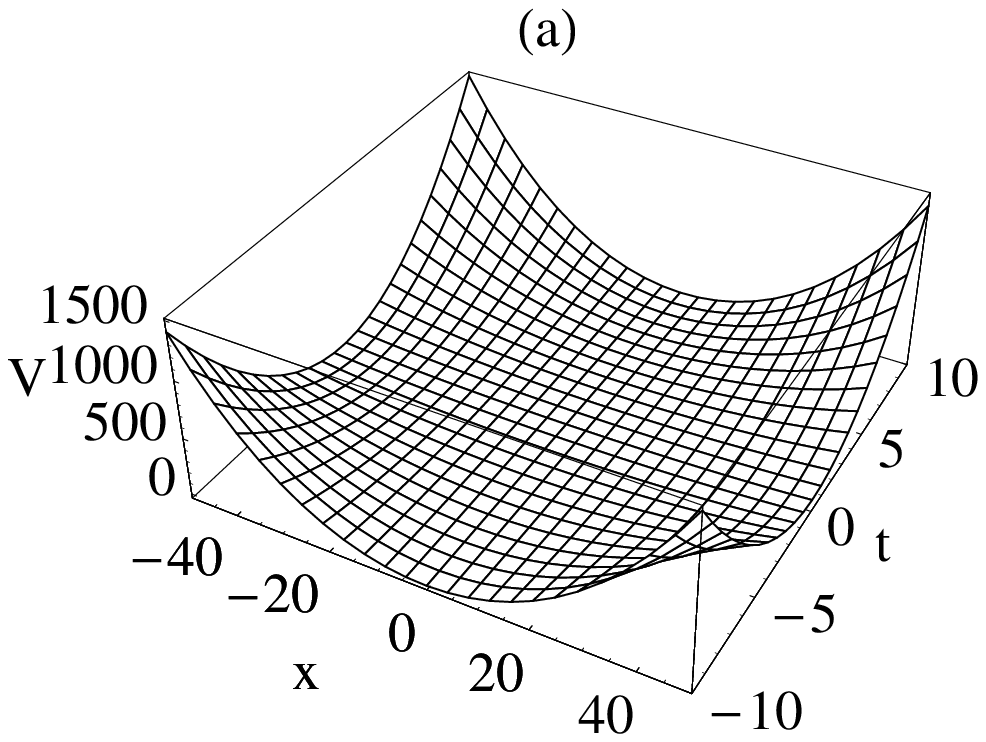}\includegraphics[scale=0.4]{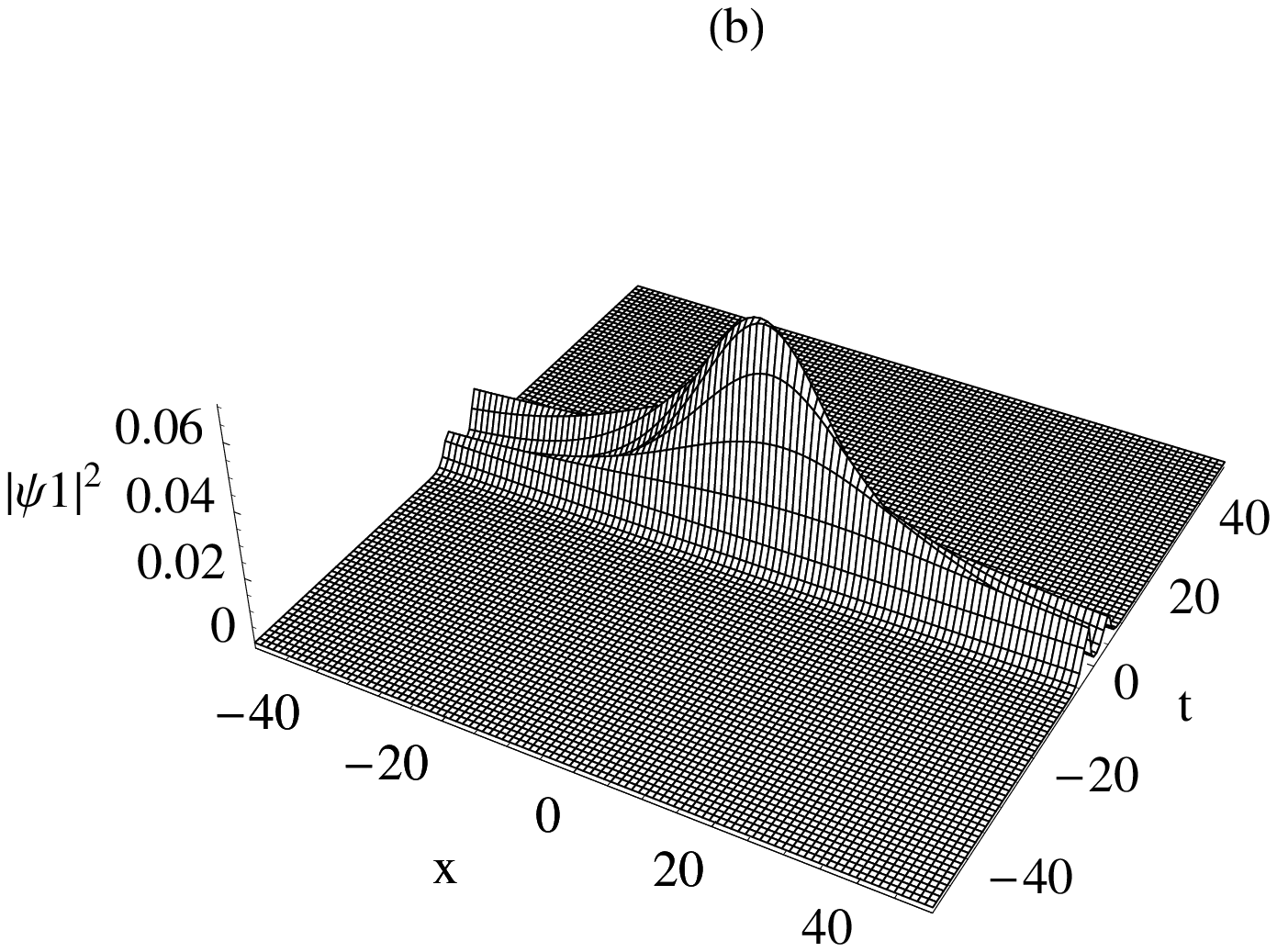}\includegraphics[scale=0.4]{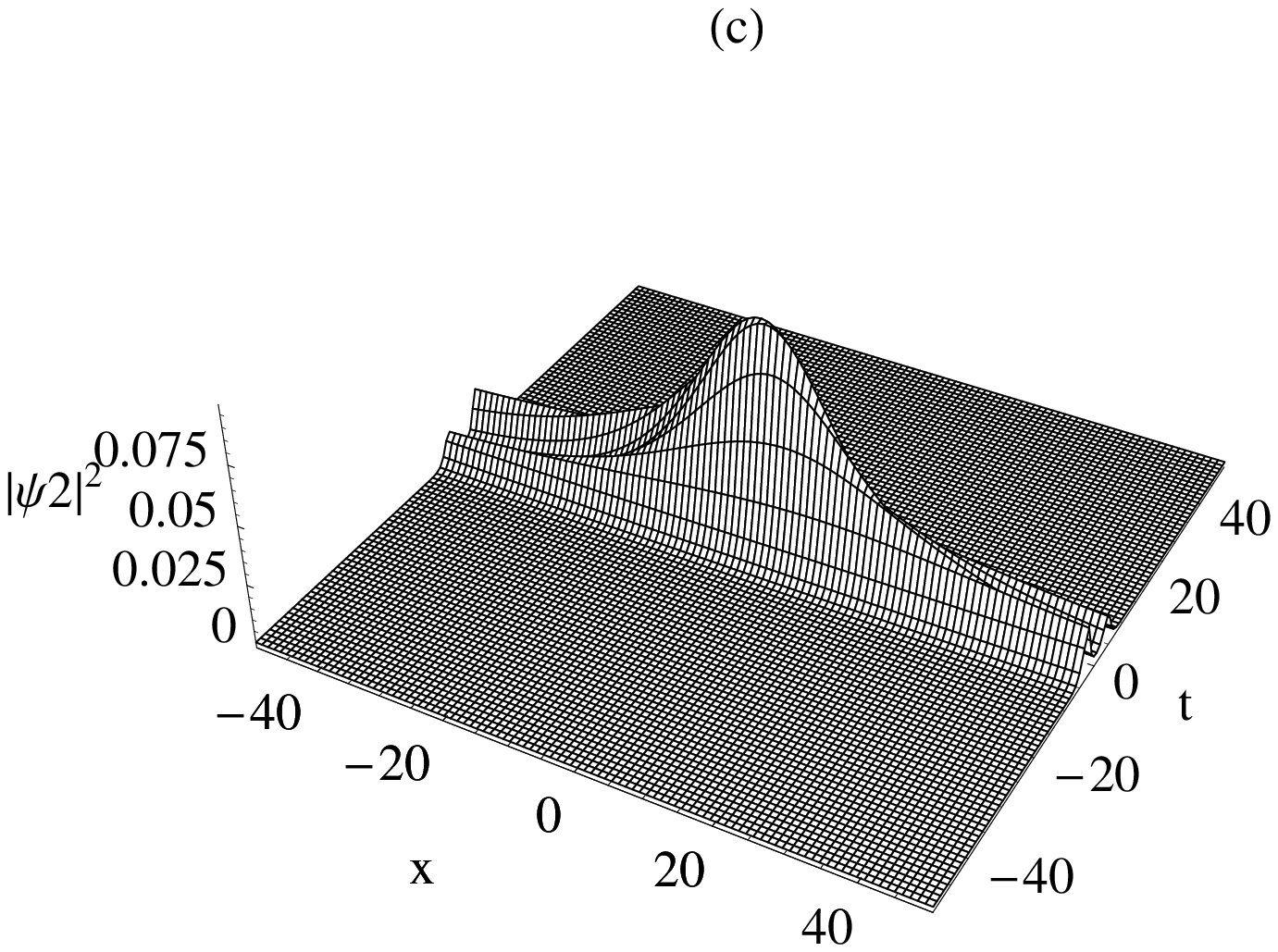}
\caption{(a) Harmonic trap for the choice $k(x)=\frac{1}{0.9 +
cos(0.02 x)^{2}}$ with $ A_1 (t)=A_2 (t)=1,{b_0 (t)}=0.1t,{b_1
(t)}=-0.2t,\theta_1(t)=\theta_2(t)=sin(0.1 t)$. (b),(c) Double
hump matter wave solitons in the harmonic trap.}
\end{figure}

\begin{figure}
\includegraphics[scale=0.4]{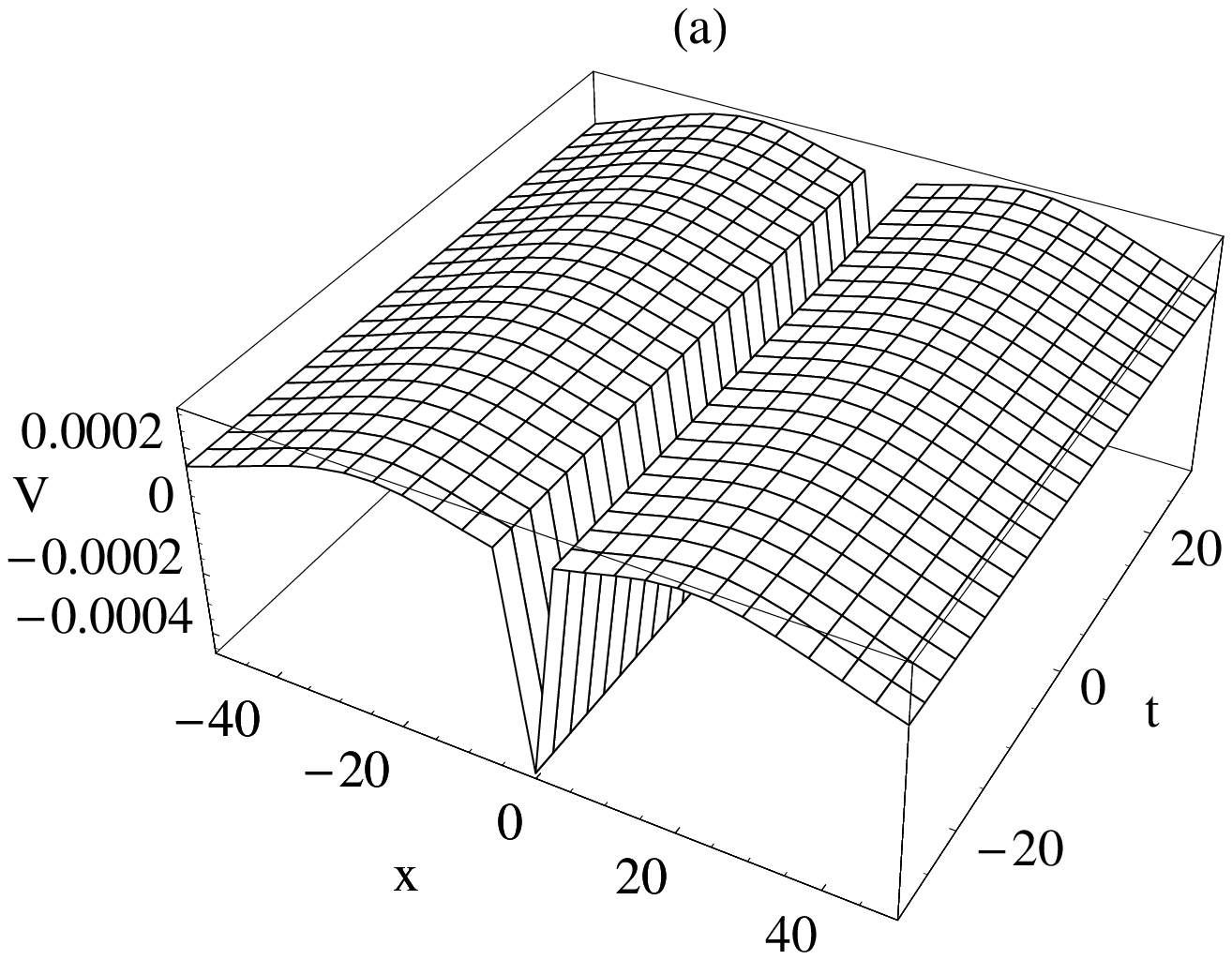}\includegraphics[scale=0.4]{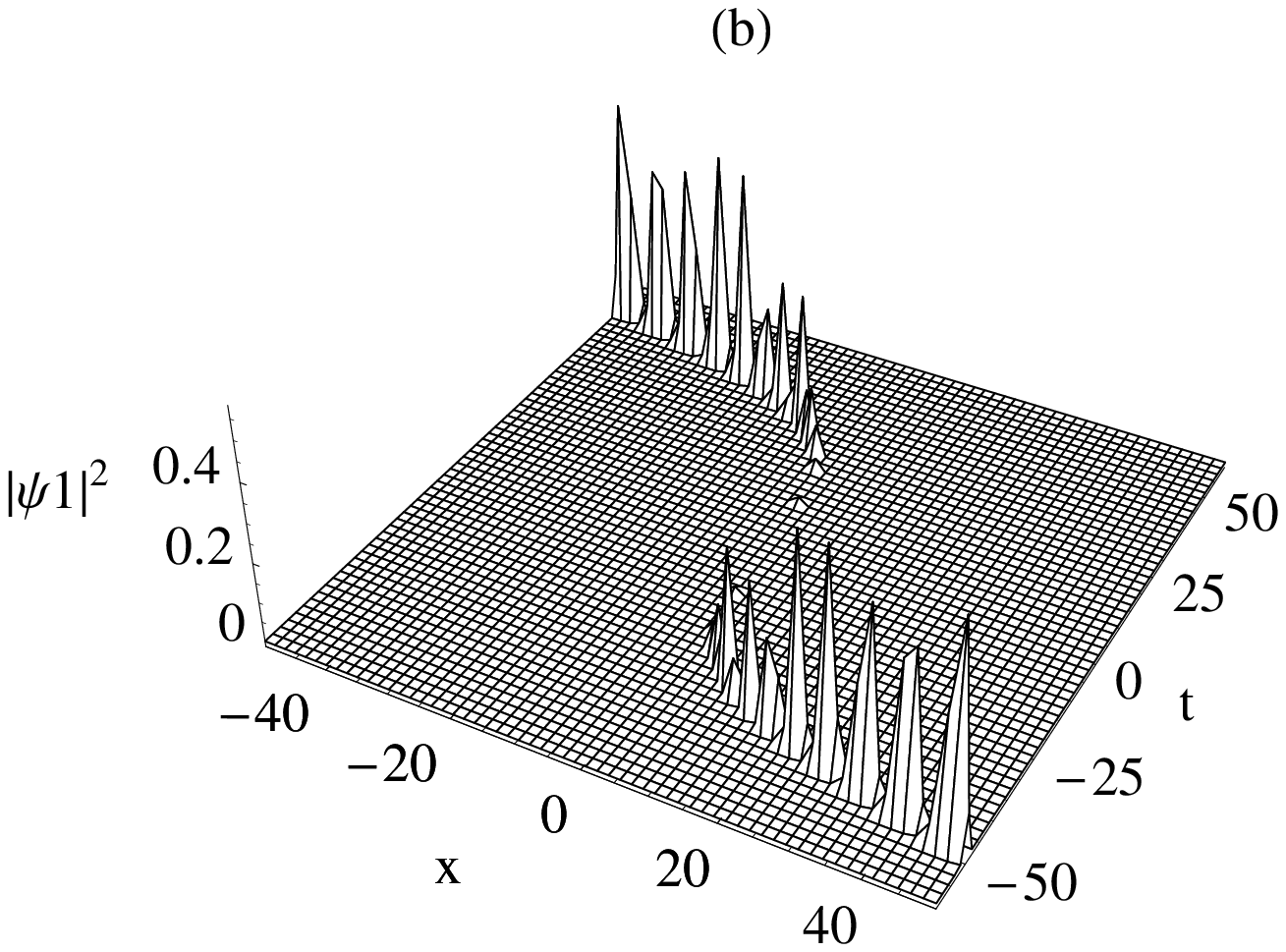}\includegraphics[scale=0.4]{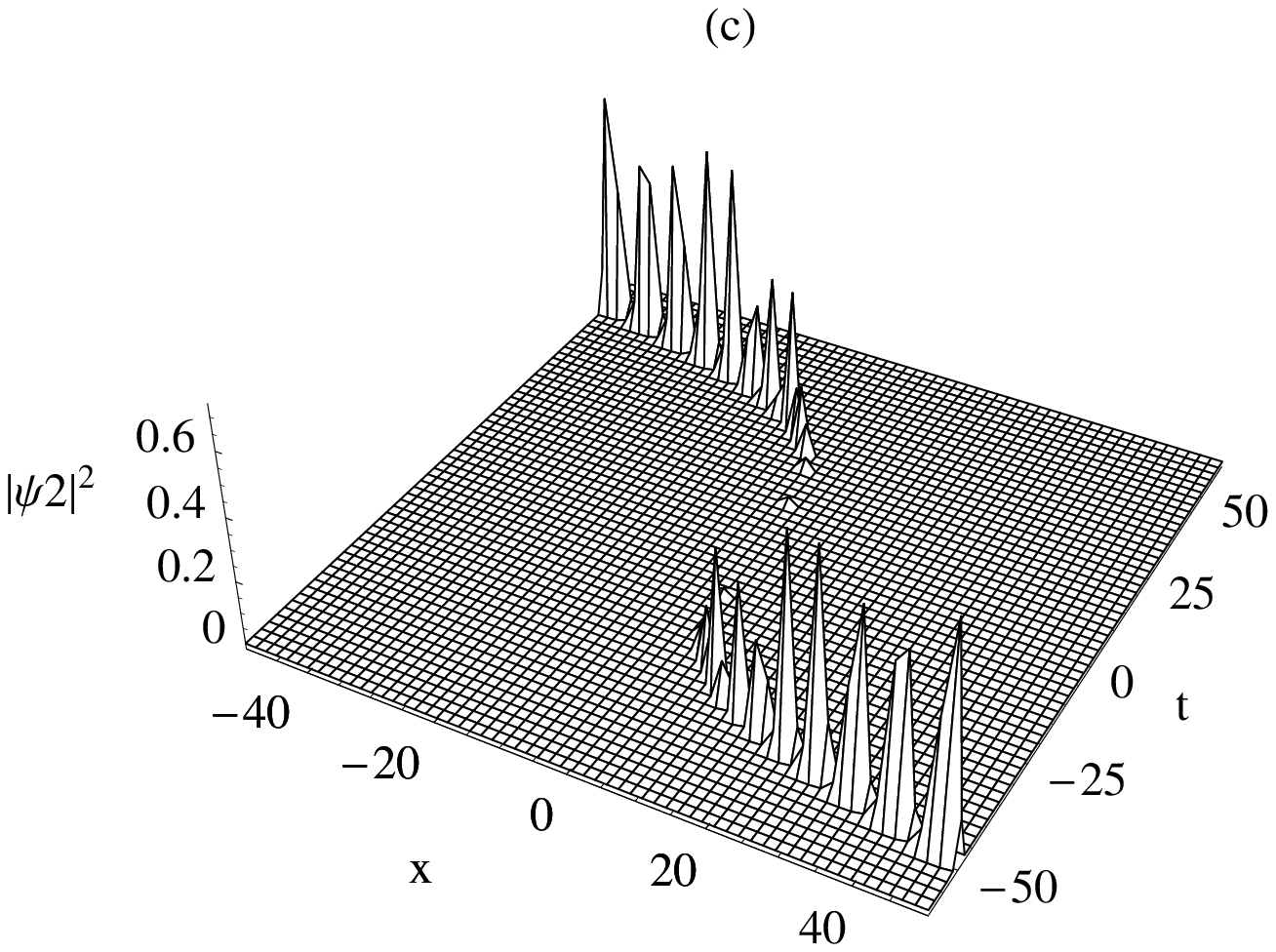}
\caption{(a)Transient trap for the  choice $k(x)=tanh(0.03 x)^{2}
+ 0.0001$ with $A_1 (t)=A_2 (t)=1$ and ${b_0 (t)}= {b_1
(t)}=\theta _1(t)=\theta _2(t)=0$,(b)(c) Growth of matter wave
solitons in the transient trap}
\end{figure}

\begin{figure}
\includegraphics[scale=0.65]{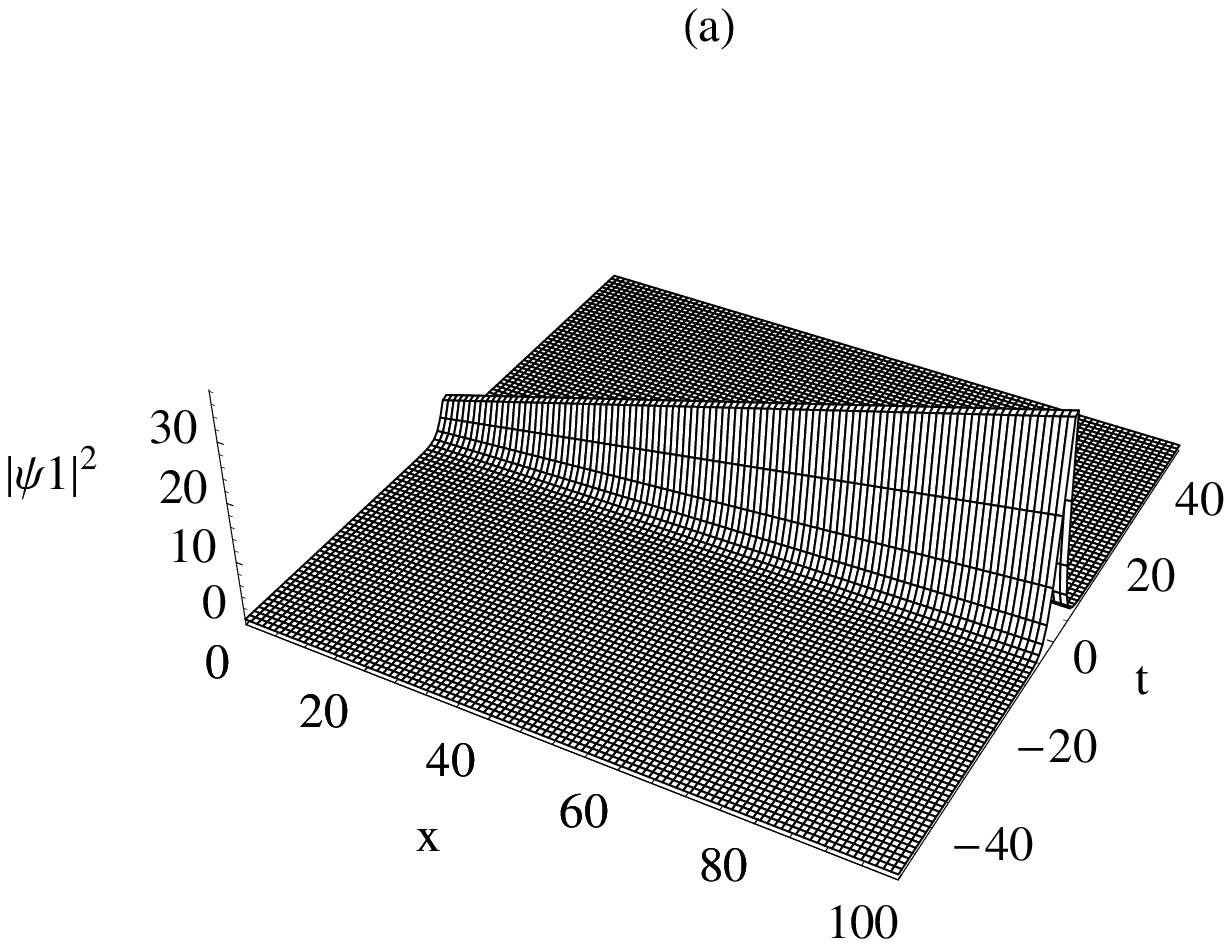}\includegraphics[scale=0.65]{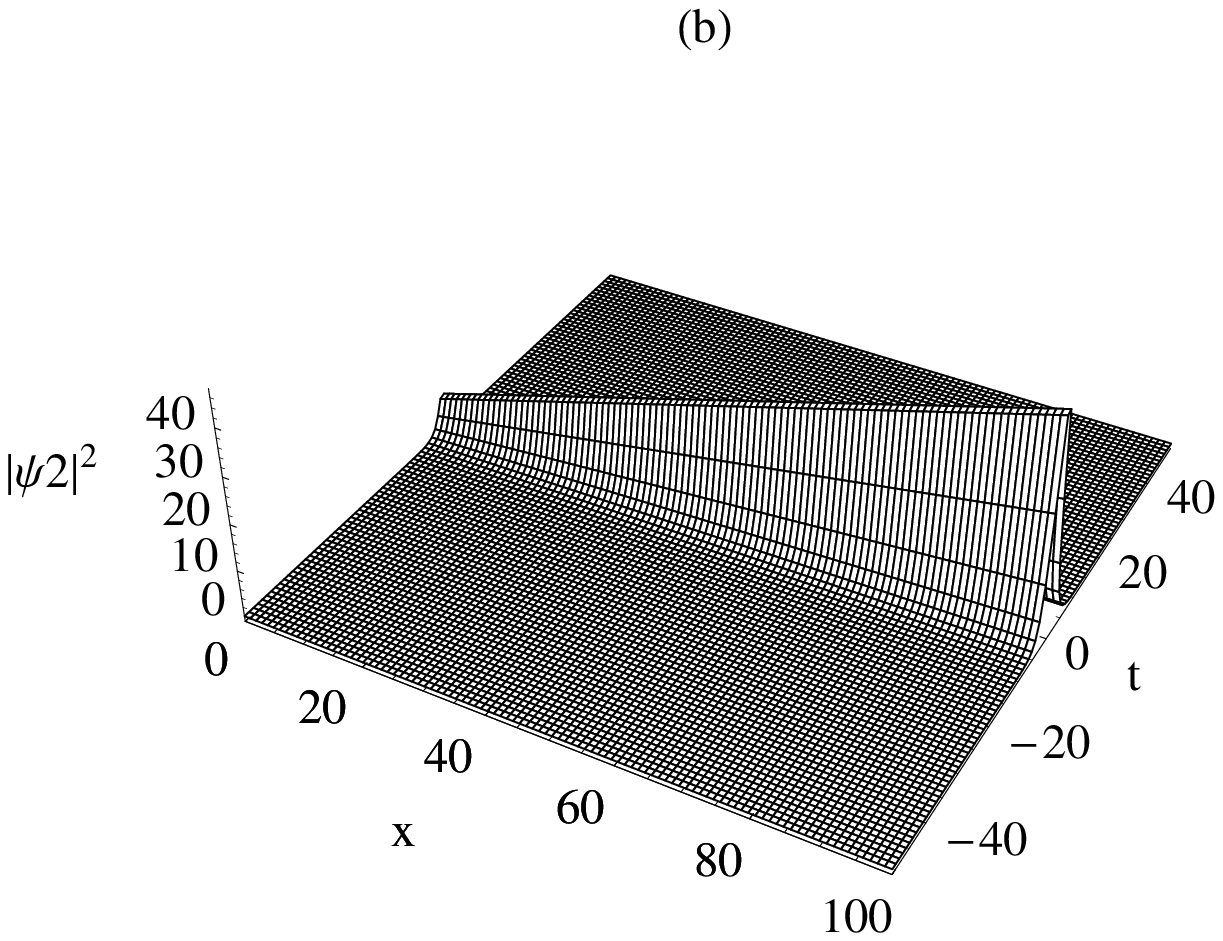}
\caption{(a),(b)Matter wave solitons in the absence of the trap
for $k(x)=(-x-4*5)^{4}$ x $0.3$}
\end{figure}

\begin{figure}
\includegraphics[scale=0.65]{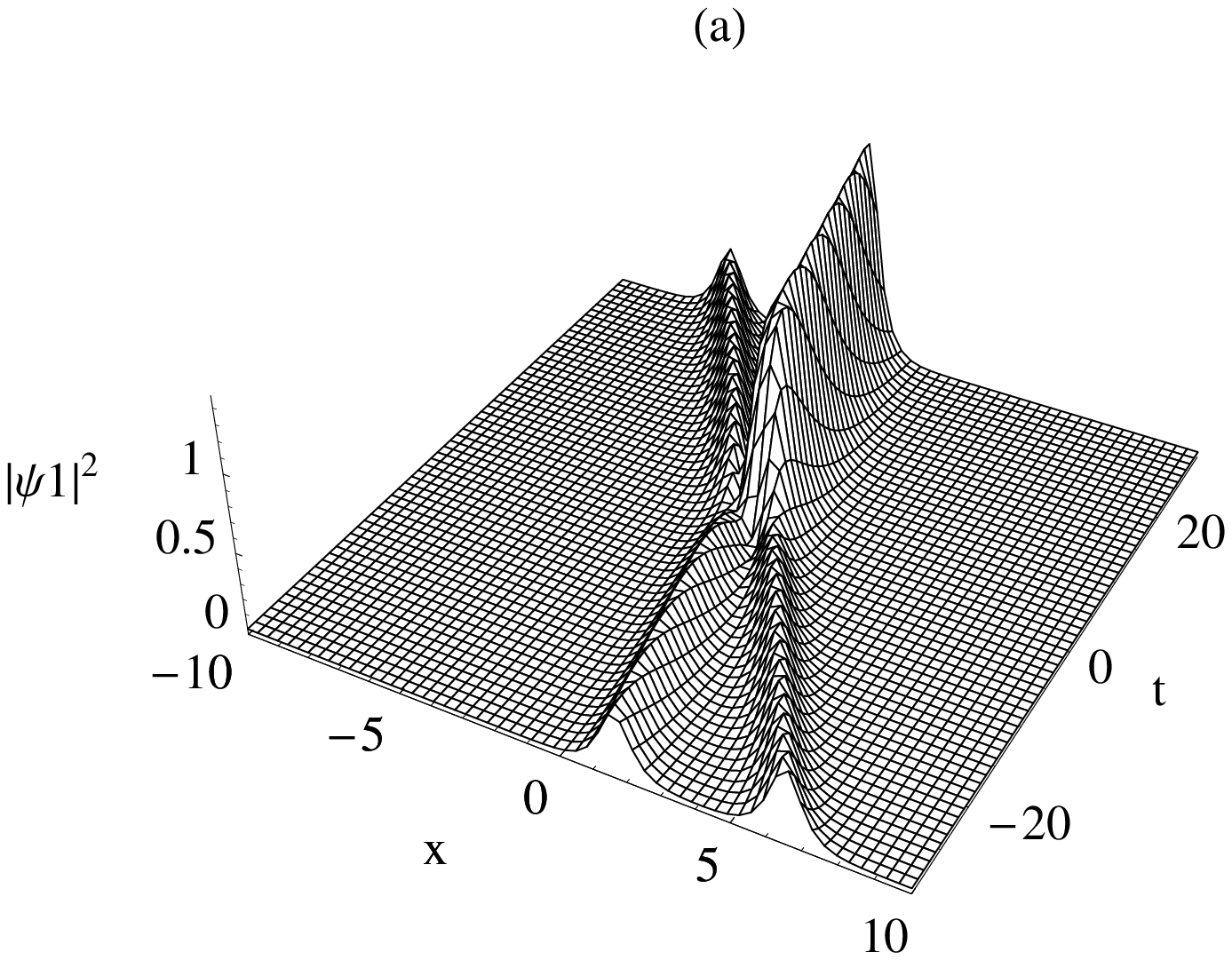}\includegraphics[scale=0.65]{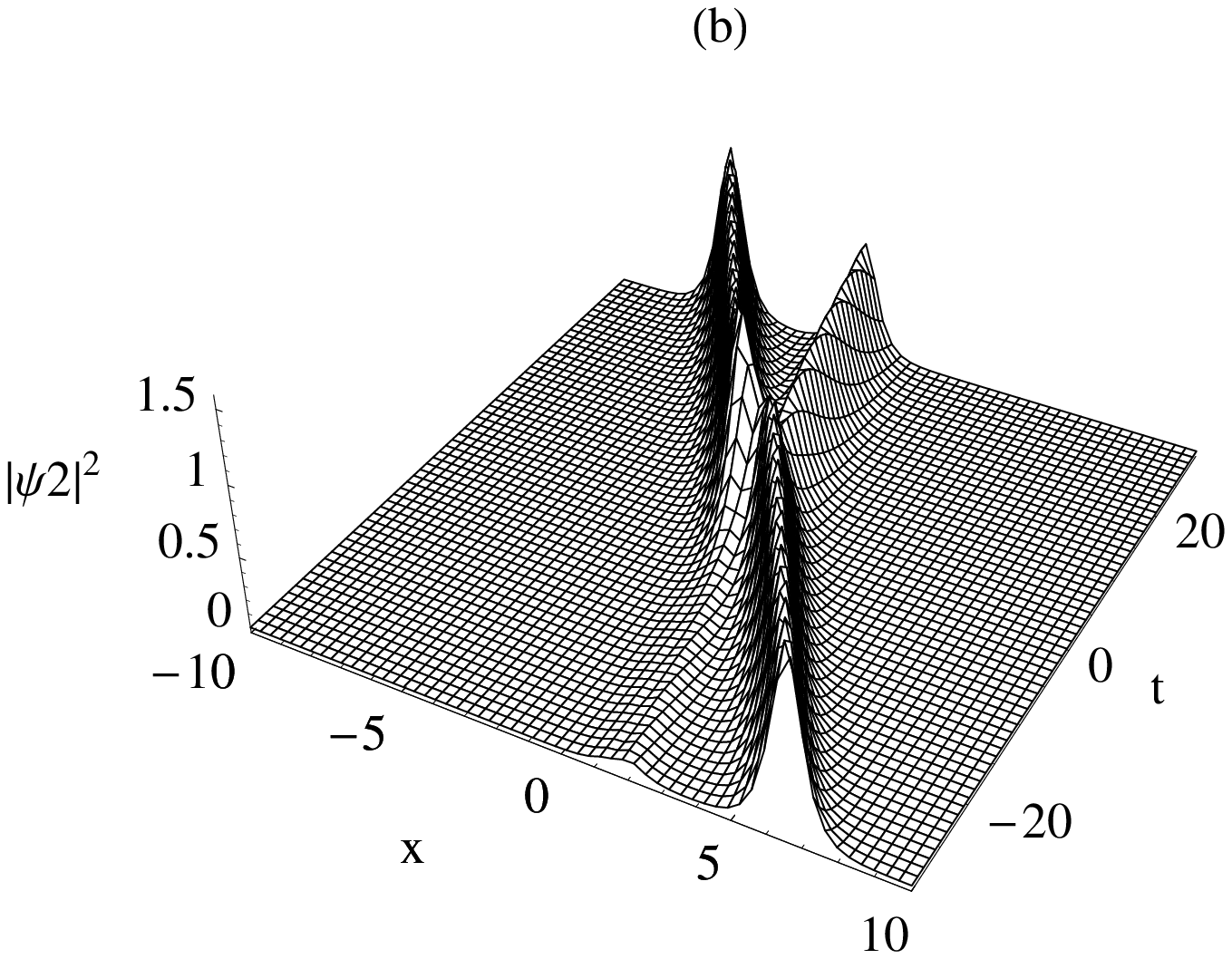}
\caption{(a)(b)Switching of matter wave solitons for
$k(x)=0.1cos(-0.3x)^{2}+ 0.1 sin(-0.3 x)^{2}$.}
\end{figure}

\begin{figure}
\includegraphics[scale=0.65]{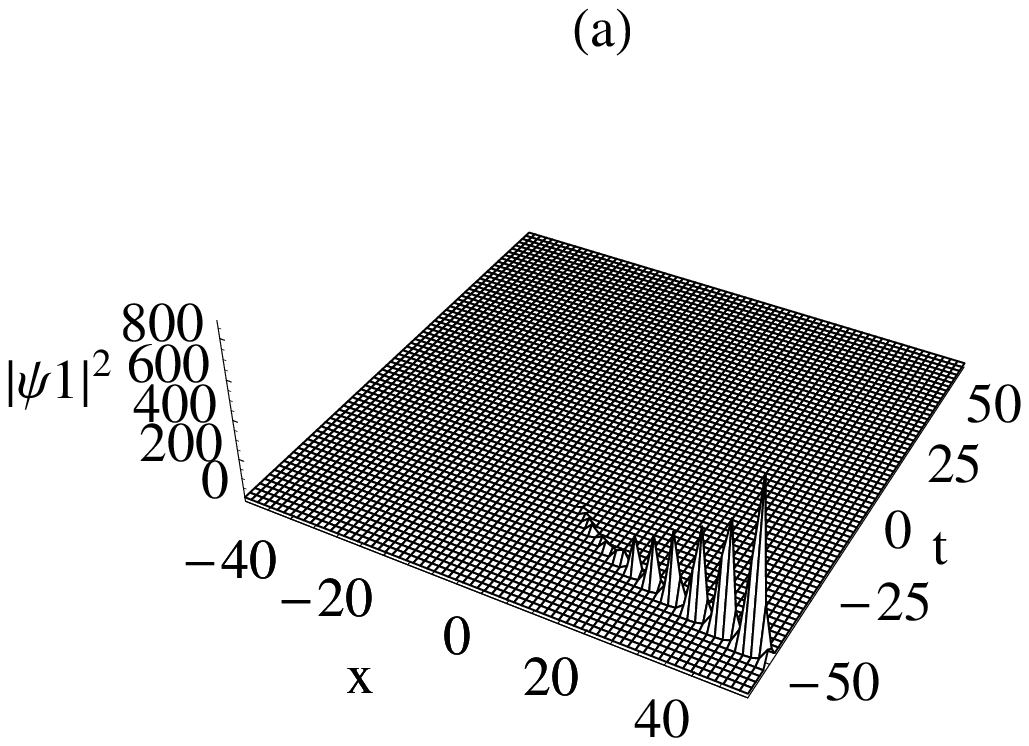}\includegraphics[scale=0.65]{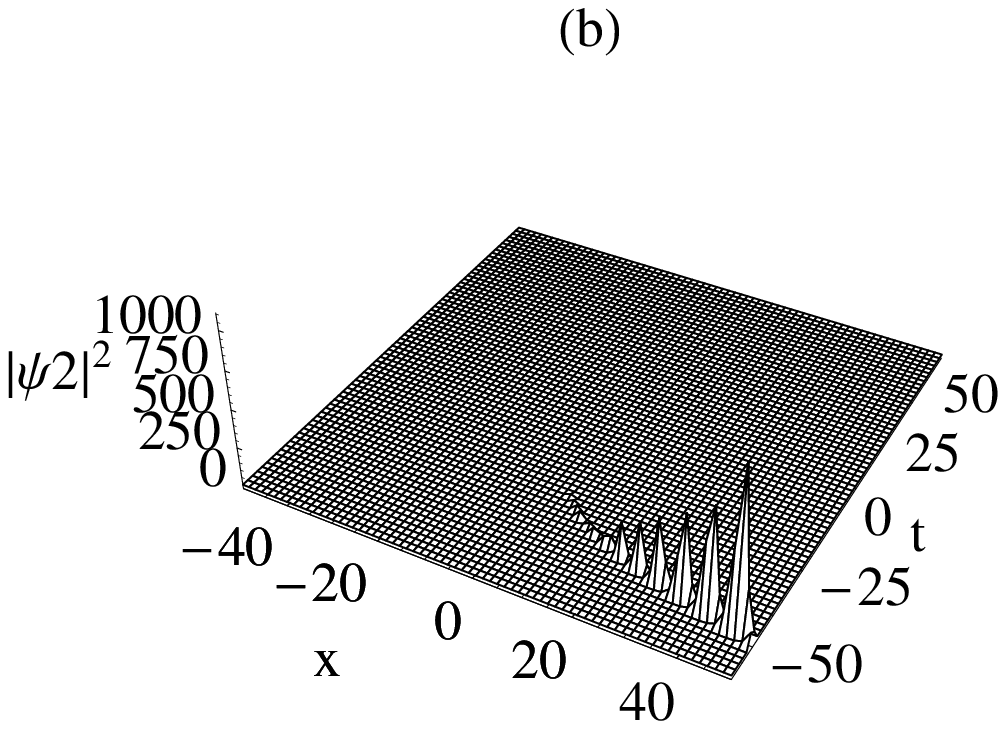}
\caption{(a,b) Impact of spatio temporal interaction and abrupt
increase of density of the condensates for $k(x)=tanh(0.03 x)^{2}
+ 0.0001$ with $A_1 (t)=A_2 (t) =e^{0.15 t},{b_0 (t)}={b_1 (t)}=0
$ and $ \theta _1(t)=\theta _2(t)=1 $ }
\end{figure}

\begin{figure}
\includegraphics[scale=0.65]{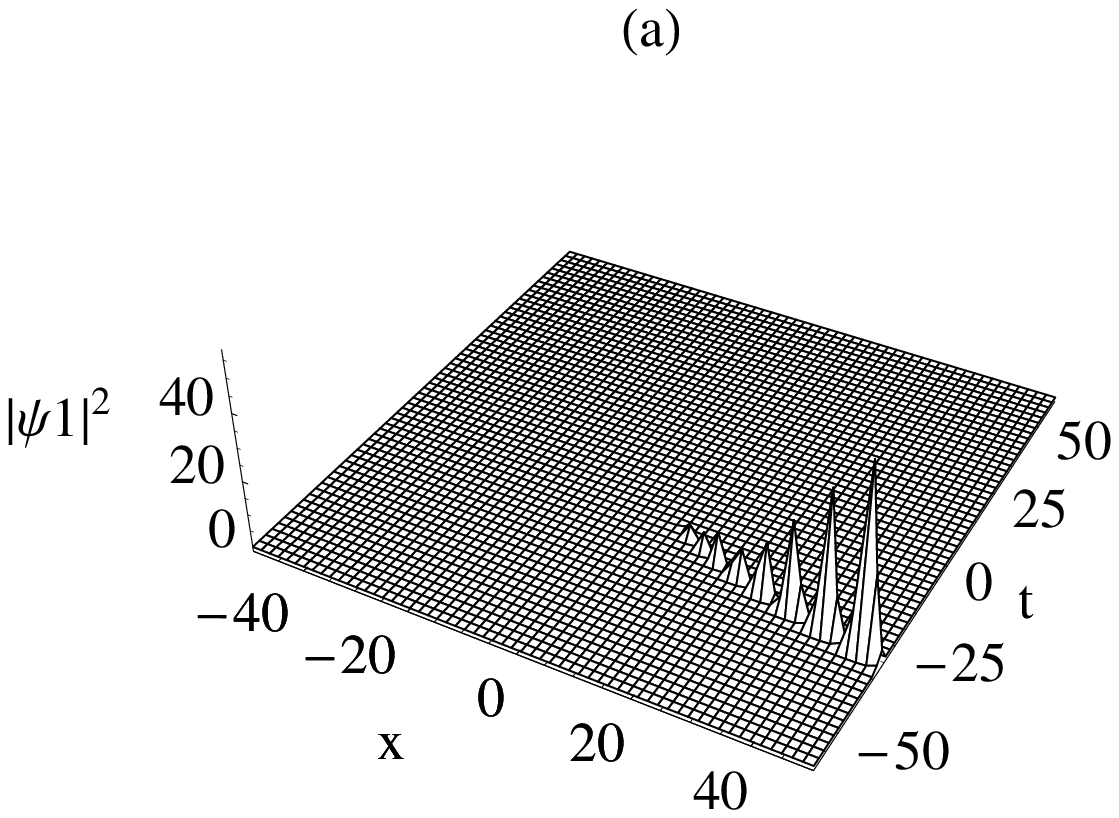}\includegraphics[scale=0.65]{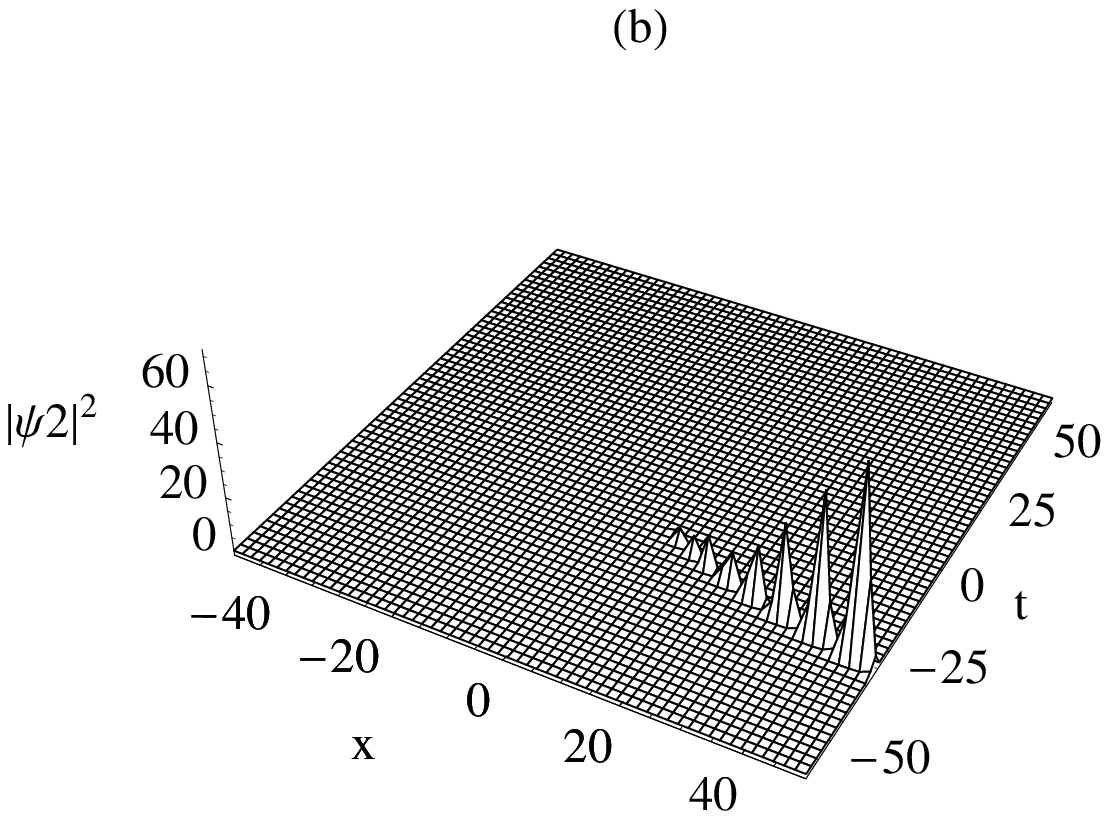}
\caption{(a,b) Stabilization of the condensates by tuning spatial
scattering length for $k(x)=tanh(0.03 x)^{2} + 0.0075 $ with $A_1
(t)=A_2 (t) =e^{0.15 t},{b_0 (t)}={b_1 (t)}=0 $ and $ \theta
_1(t)=\theta _2(t)=1 $ }
\end{figure}

\subsection{Matter wave solitons in harmonic traps}
Selecting $k(x)=\frac{1}{0.9 + cos(0.02 x)^{2}}$ and again
choosing the parameters $\theta_1 (t), \theta_2 (t), {b_0 (t)}$
and $ {b_1 (t)}$ desirably, one obtains a harmonic trap as shown
in figure (2a). Accordingly, one observes double hump solitons for
the condensates $|\psi_1|^{2}$ and $|\psi_2|^{2}$ in figures (2b)
and (2c) respectively. Again, the width and amplitude of the
bright solitons can be altered by changing the parameters
associated with $k(x)$. This underscores the fact that the energy
associated with the modes $|\psi_1|^{2}$ and $|\psi_2|^{2}$ can be
modulated desirably.

\subsection{Transient trap behaviour} Choosing the transient
trap as shown in fig (3a) (after appropriately choosing $k(x)$ and
the parameters $\theta_1 (t), \theta_2 (t), {b_0 (t)}$ and $ {b_1
(t)}$ suitably), the behaviour of the condensates in the transient
trap is shown in figures (3b) and (3c). From the figures 3(a-c),
one could observe the stabilization of the condensates in the
confining trap  while the life time of the expulsive domain is too
short to notice the dynamics of the condensates.

\subsection{Switching off the trap and bright soliton dynamics}
Choosing the dispersion coefficient $k(x)= (-x-4$ x $5)^4$ x
$0.3$, we observe that the external trap $V(x)$ becomes zero.
Under this condition, we observe that the amplitude of the
solitons keeps growing as they evolve in space and time as shown
in figs (4a,b)and this increase of amplitude of matter wave
solitons occurs without any addition of external energy for a
specific choice of spatially inhomogeneous interaction.

$\quad$The gauge transformation approach can be easily extended to
generate multisoliton solution [33] and the interaction of
solitons can be analysed.

\subsection{Matter wave switching} Choosing the spatially
varying dispersion coefficient of the form $k(x)=0.1cos(-0.3
x)^{2}+ 0.1 sin(-0.3 x)^{2}$, one observes switching of
intensities of matter wave solitons as shown in the figs (5a,b).
Thus, it is obvious from the figs (5 a,b) that the intensity
redistribution occurs between the modes $|\psi_1|^{2}$ and
$|\psi_2|^{2}$.

\subsection{Impact of spatio temporal interaction on the condensates} It is obvious from the above that cases [(i)-(v)]
discuss the effect of spatially inhomogeneous interaction alone on
the condensates. To investigate the impact of spatiotemporal
interaction on the condensates, we now consider the evolution of
the condensates in the transient trap shown in figure (3a)
corresponding to case (iii). Now, evolving the temporal
 scattering lengths of the form $A_1 (t)=A_2 (t)
 =e^{0.15 t} $,the density of the condensates which were stabilized in the
confining domain in the absence of temporal interaction now
abruptly increases as shown in the figs (6a,b). We now suitably
tune the spatially inhomogeneous interaction which subsequently
reduces the density of the condensates. Hence, we observe that it
should be possible to vary the spatially inhomogeneous interaction
to stabilize the condensates just as one temporally varies the
scattering length using Feshbach resonance as displayed in figs
(7a,b) . It should be mentioned that this is the first instance of
the occurrence of Feshbach resonance employing the variation of
spatially inhomogeneous interaction. The combined effect of
spatiotemporal interaction is that one can not only control the
density of the condensates, but also design condensates with
desirable density, shape (or geometry) and property leading to an
era of \textbf{"designer quasi particle condensates"}.
\section{Discussion}
In this paper, we  have derived a new integrable model to
investigate the dynamics of two component quasi-particle
condensates with spatiotemporal interaction strengths and
construct the associated Lax pair. We then generate the matter
wave solitons and study their properties in harmonic and optical
lattice potentials.

We also report the occurrence of Feshbach resonance by subtle
variation of spatially inhomogeneous interaction. We reiterate
that the simultaneous impact of spatio temporal interaction could
possibly herald a new era of "\textbf{designer quasi particle
condensates".}

\section{Acknowledgements} PSV wishes to thank UGC and DAE-NBHM for financial support.
The work of RR forms part of a research project sponsored by
DST,DAE -NBHM and UGC. Authors thank the anonymous referee for his
suggestions. RR wishes to thank Prof. Malomed for giving
invaluable suggestions. KP acknowledges DST and CSIR, Government
of India, for the financial support through major projects.


\newpage
\section*{References}

\begin{thebibliography}{26}
\bibitem{ref1}
Anderson M H, Ensher J R, Matthews M R, Wieman C E and Cornell E A
1995 $Science$ \textbf{269} 198
\bibitem{ref2}
Davis K B, Mewes M O, Andrews M R, VanDruten N J, Durfee D S, Kurn
D M and Ketterle W 1995 $Phys. Rev. Lett$ \textbf{75} 3969
\bibitem{ref3}
Dalfovo F, Giorgini S, Pitaevskii L P and Stringari S 1999 $Rev.
Mod.Phys$ \textbf{71} 463
\bibitem{ref4}
Gross E P 1961 $Nuovo$ $Cimento$ \textbf{20} 454
\bibitem{ref5}
Gross E P 1963 $J. Math. Phys$ \textbf{4} 195
\bibitem{ref6}
Pitaevskii L P 1961 $Zh.Eksp.Teor.Fiz$ \textbf{40} 646 $[$J. Exp.
Theor. Phys $\textbf{13}$ 1961 451$]$
\bibitem{ref7}
Liang Z X, Zhang  Z D and Liu  W M 2005 $Phys. Rev. Lett$
\textbf{94} 050402
\bibitem{ref8}Radha R and Ramesh Kumar V 2007 $Phys. Lett. A$
\textbf{370} 46
\bibitem{ref9}
Radha R, Ramesh Kumar V and Porzeian K 2008 $J. Phys. A: Math.
Theor$ \textbf{41} 315209
\bibitem{ref10}
Ramesh Kumar V, Radha R and Panigrahi P K 2008 $Phys. Rev. A$
\textbf{77} 023611
\bibitem{ref11}
Strecker K E, Partridge G B, Truscott A G and Hulet R G  2002
$Nature$ \textbf{417} 150
\bibitem{ref12}
Khaykovich L, Schreck F, Ferrari G, Bourdel  T, Cubizolles  J,
Carr L D, Castin  Y, Solomon  C 2002 $Science$ \textbf{296} 1290
\bibitem{ref13}
Strecker K E, Partridge G B, Truscott  A G and Hulet R G 2003
$New.J.Phys$ \textbf{5} 73
\bibitem{ref14}
Burger S, Bongs K, Dettmer S, Ertmer W, Sengstock K, Sanpera A,
Shlyapnikov G V and Lewenstein M 1999 $Phys. Rev. Lett$
\textbf{83} 5198
\bibitem{ref15}
Denschlag J, Simsarian J E, Feder D L, Clark C W, Collins  L A,
Cubizolles J, Deng L, Hagley E W, Helmerson K, Reinhardt W P,
Rolston S L, Schneider B I, Phillips W P 2000 $Science$
\textbf{287} 5450.97.
\bibitem{ref16}
Papp S B, Pino  J M and Wieman  C E 2008 $Phys. Rev. Lett$
\textbf{101} 040402
\bibitem{ref17}
Thalhammer G, Barontini  G, Sarlo  L De, Catani J, Minardi  F and
Inguscio M 2008 $Phys. Rev. Lett$ \textbf{100} 210402
\bibitem{ref18}
Nathan Kutz J 2009 $Physica D$ \textbf{238}  1468
\bibitem{ref19}
MiddelKamp S, Chang  J J, Hamner  C, Carretero-Gonzalez  R,
Kevrekidis P G, Achilleos  V, Frantzeskakis  D J, Schmelcher  P,
Engels P 2010 $Phys.Lett.A$ \textbf{375} 642
\bibitem{ref20}
Theocharis G, Schmelcher P, Kevrekidis P G, Frantzeskakis D J 2005
$Phys. Rev.A $ \textbf{72} 033614
\bibitem{ref21}
Rajendran S, Muruganandam P and Lakshmanan  M 2009 $J.Phys.B. At
Mol.Opt.Phys$ \textbf{42} 145307
\bibitem{ref22}
Ramesh Kumar  V, Radha R and Wadati M 2010 $Phys. Lett.A$
\textbf{374} 3685
\bibitem{ref23}
Rodas-Verde M I, Michinel H and Perez-Garcia  V M 2005 $Phys. Rev.
Lett$ \textbf{95} 153903
\bibitem{ref24}
Carpentier A V, Michinel  H, Rodas-Verde M I and Perez-Garcia V M
2006 $Phys. Rev. A$ \textbf{74} 013619
\bibitem{ref25}
Shin H J, Radha  R, Ramesh Kumar V 2011 $Phys. Lett.A$ \textbf{
375} 2519
\bibitem{ref26}
He J S, Mei J and Li Y S 2007 $Chin. Phys. Lett.$ \textbf{24} 2157
\bibitem{ref27}
He J S and Li Y S 2011 $Stud. Appl. Math.$ \textbf{126} 1
\bibitem{ref28}
Wang Y Y, He J S and Li Y S 2011 $Commun. Theor. Phys.$
\textbf{56} 995
\bibitem{ref29}
Xu S W, He J S and Wang L H 2012 $Europhys. Lett.$ \textbf{97}
30007
\bibitem{ref30}
He X G, Zhao  D, Li  L and Luo  H G 2009 $Phys. Rev. E$
\textbf{79} 056610
\bibitem{ref31}
Wen L, Li L, Li Z D, Song S W, Zhang X F and Liu W M 2011 $Eur.
Phys. J. D$ \textbf{64} 473
\bibitem{ref32}
Manakov S V 1974 $Sov.Phys.JETP.$ \textbf{38} 248
\bibitem{ref32}
Chau L L, Shaw J C and Yen H C 1991 $J. Math. Phys$ \textbf{32}
1737
\bibitem{ref33}
lizuka T, Wadati M 1997 $J. Phys. Soc. Jpn$ \textbf{66} 2308

\end{thebibliography}
\end{document}